\documentclass[journal]{IEEEtran}
\usepackage{cite}
\usepackage{graphicx}
\usepackage{amsmath}
\usepackage{amsfonts}
\usepackage{array}
\usepackage{amsthm}
\usepackage{amssymb}
\usepackage{mathtools}
\usepackage{epstopdf}
\usepackage{bm}
\begin{document}

	\title{Characterizing Inter-Numerology Interference in Mixed-Numerology OFDM Systems }
	\author{Juquan Mao,~Lei Zhang, Stephen McWade,  Hongzhi Chen, Pei Xiao} 
		%\thanks{J. Mao, H. Chen and P. Xiao are with the Institute for Communication Systems (ICS), Home of 5G Innovation Centre, University of Surrey, Guildford, GU2 7XH, UK (e-mail:\{juquan.mao, hongzhi.chen,\}@surrey.ac.uk).}

	\maketitle

	\begin{abstract}
		
		The advent of mixed-numerology multi-carrier (MN-MC) techniques adds flexibilities in supporting heterogeneous services in fifth generation (5G) communication systems and beyond. However, the coexistence of mixed numerologies destroys the orthogonality principle defined for single-numerology orthogonal frequency division multiplexing (SN-OFDM) systems with overlapping subcarriers of uniform subcarrier spacing. Consequently, the loss of orthogonality leads to inter-numerology interference (INI),  which complicates signal generation and impedes signal isolation. In this paper, the INI in MN-OFDM systems is characterized through mathematical modeling and is shown to primarily rely on system parameters with regard to the pulse shape, the relative distance between subcarriers and the numerology scaling factor. Reduced-form formulas for the INI in continuous-time and discrete-time MN systems are derived. The derived mathematical framework facilitates the study of the effect of discretization on the INI and partial orthogonality existing in subsets of the subcarriers. The reduced-form formulas can also be used in developing interference metrics and designing mitigation techniques.
		
	\end{abstract}


	\IEEEpeerreviewmaketitle

	\section{Introduction}\label{sec:introduction}

		5G and beyond mobile networks are envisioned to have the  flexibility to support heterogeneous services.  These services have been broadly categorized into three main usage scenarios: enhanced mobile broadband (eMBB), ultra-reliable and low-latency communications (uRLLC), and massive machine type communications (mMTC) \cite{series2015imt}. Each of these scenarios has distinct quality of service (QoS) requirements, such as throughput, latency, reliability, and the number of connected users, which calls for a higher degree of flexibility in the physical layer network designs\cite{yang2020mixed}. It is obvious that a one-size-fits-all numerology design, as in 4G long term evolution (LTE), may not be able to provide the desired flexibility. In addition, it is not  viable to design separate radios for different services due to the significantly increased complexity of operation and management \cite{zhang2017multi}.

		To support these services over a unified physical layer, one solution is to divide the system bandwidth into several smaller bandwidth parts (BWPs), each having a distinct numerology (a set of parameters like subcarrier spacing, symbol length, and cyclic prefix \cite{zaidi2016waveform}) optimized for a particular service.  However, allowing the coexistence of mixed numerologies in the same carrier introduces non-orthogonality to the system and causes inter-numerology interference (INI) because subcarriers associated with differing numerologies will no longer be orthogonal to each other. The loss of orthogonality complicates both signal generation and detection in MN-OFDM systems.

		The study of  INI  has recently attracted increasing interest. The authors of \cite{zhang2017subband} developed a framework for MN subband filtered multi-carrier (SFMC) systems where the INI was analyzed in the presence of transceiver imperfections and insufficient guard interval between symbols. The authors of \cite{yli2017efficient} introduced a generic and universal optimization-based framework for fast convolution-based filtered OFDM (FC-f-OFDM) waveform processing for the 5G physical layer in which the INI for MN FC-f-OFDM was derived. The authors of \cite{zhang2018mixed} derived expressions for the INI  in MN windowed OFDM systems and used them as the basis for a novel interference cancellation scheme. An analytical  model was established in \cite{mao2020interference}  for INI analysis in OFDM and f-OFDM systems,  and a power allocation scheme was also proposed to maximize the system sum-rate utilizing the derived analysis. 

		All the works above have derived closed-form formulas for INI for various MN systems based on a finite-sum of discrete samples of complex exponential functions, referring to as sum-of-exponential (SoE) forms in the sequel. However, these SoE formulas can not explicitly show the influencing factors on the interference.  In this paper, we progressed further by simplifying the SoE to a reduced-form of INI to show the influencing factors more explicitly.  Moreover, all existing works have focused on discrete-time systems; to the researchers' best knowledge, this is the first time investigating continuous-time MN systems. Specifically, this paper contributes to the following:
		\begin{itemize}
			\item We first derive analytical expressions for the INI in continuous-time systems. This work is essential as continuous-time approaches have some notable advantages in performance degradation analysis due to system imperfections, such as time/frequency offset and phase noise,  compared with discrete-time methods.
			\item We simplify the traditional SoE form INI into a reduced-form, which is primarily a pulse shape function of relative distance. i.e., $\text{sinc}(d)$, where $d$ refers to the relative distance between MN subcarriers. The reduced-form not only decreases the computational complexity but also directly shows the factors influencing the level of INI.
			\item We study the impact of discretization on the INI and discover the fact that the discrete-time subcarriers suffer more interference than the continuous ones.
			\item We investigate the subsets of MN subcarriers based on the derived reduced-form formulas for both continuous-time and discrete-time systems.
		\end{itemize}

		The rest of this paper is organized as follows: Section \ref{sec:system_model} describes the MN-OFDM system model. Section \ref{sec:cha_INI} characterizes the INI in the continuous-time and discrete-time MN systems. Finally, the paper is concluded in Section \ref{sec:con}. Throughout this paper, $(\cdot)^*$ denotes complex transpose,  $\mathbf R$ and $\mathbf Z$ represents the set of all real numbers and integers, respectively, and the superscript $^{(i)}$ is used to denote numerology $i$. Often we deal with functions of a continuous variable, and a related sequence indexed by an integer (typically, the latter is a sampled version of the former). We use parentheses around a continuous variable and brackets around a discrete one, for example, $x(t)$ and $x[n]$. 

	\section{Signal Model}\label{sec:system_model}

		\begin{figure}[t]
			\centering
			\includegraphics[width=1\linewidth]{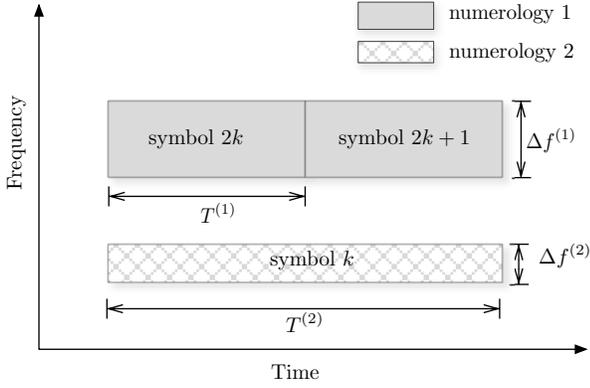}
			\caption{Illustration of frequency and time relationship with two numerologies for the case $\nu =2$. }
			\label{fig:timefreqillustration}
		\end{figure}

		For clarity and simplicity of the derivations, without loss of generality, we consider a signal model with two differing numerologies, namely, numerology $1$ and numerology $2$\footnote{Research findings gained in this paper may extend to any number of numerologies in a straightforward manner.}. The subcarrier spacing $\Delta f^{(i)}$ and symbol duration $T^{(i)}$ associated  with numerology $i$, $i=\{1,2\}$,  are related to each other via a scaling factor $\nu$, as per 3GPP \cite{3gpp_38_211}, i.e.,
		\begin{align} \label{eq:scaling}
			\frac{\Delta f^{(1)}}{\Delta f^{(2)}} &= \frac{T^{(2)}}{T^{(1)}}= \nu.
		\end{align}

		Without loss of generality, we assume ${\Delta f^{(1)}} > {\Delta f^{(2)}}$, which leads to the scaling factor greater than 1, i.e., $\nu= 2^{\mu}, \mu \in \{1,2,\cdots\} $. Fig. \ref{fig:timefreqillustration} illustrates the time/frequency relation with two numerologies for the case $\nu =2$ where $\Delta f^{(1)} = 2\Delta f^{(2)}$ and $\Delta T^{(2)} = 2\Delta T^{(1)}$, and the symbol $k$ associated with numerology 2 overlaps with two symbols indexed at $2k$ and $2k+1$ associated with numerology 1.

\begin{figure}[t]
			\centering
			\includegraphics[width=1\linewidth]{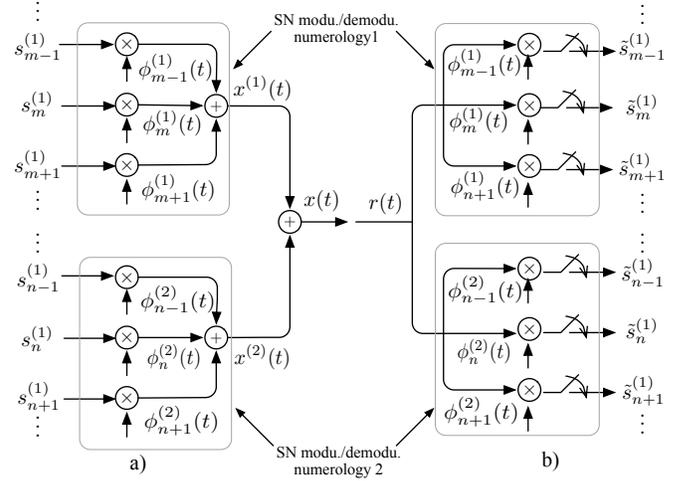}
			\caption{A signal model with two numerologies. a)  Multi-numerology OFDM modulator. b)  Multi-numerology OFDM demodulator.}
			\label{fig:systemmodel}
		\end{figure}

		Fig. \ref{fig:systemmodel} depicts a block diagram of the MN-OFDM modulator/demodulator consists of two SN-OFDM modulators/demodulators associated with the two numerologies, respectively. A group of complex symbols $\mathbf s^{(i)}$ is modulated via an SN-OFDM modulator for numerology $i$. Let the complex transmit symbol at the time instance $k$ on the subcarrier $m$ associated with numerology $i$ be $ {s^{(i)}_{k,m}}$. The transmitted signal for numerology $i$ can then be expressed as
		\begin{align}
			x^{(i)}(t) &= \sum_{k= -\infty}^{\infty} x^{(i)}_k(t),%\quad t\in [kT^{(i)}, (k+1)T^{(i)}]
		\end{align}
		where
		\begin{align}\label{eq:sn_sig}
			x^{(i)}_k(t) =\sum_{k= -\infty}^{\infty}\sum_{m= 0}^{N^{(i)}-1} {s^{(i)}_{k,m}} {\phi^{(i)}_m}(t-kT^{(i)})
		\end{align}
		is the corresponding transmitted signal at the instance $k$. $N^{(i)} = B/\Delta f^{(i)}$ is the number of subcarriers associated with numerology $i$ with $B$ denoting the overall system bandwidth.  The basis pulse $\phi^{(i)}_m(t)$ is a normalized, frequency-shifted  rectangular pulse defines as
		\begin{align}\label{eq:complex_exp}
			\phi^{(i)}_m(t) = 
			\begin{cases}
				\begin{aligned}
					&\frac {1}{\sqrt{T^{(i)}}}\exp(j2\pi\frac{m}{T^{(i)}} t)& 0\leq t \leq T^{(i)}\\
					\\&0 &\text{otherwise}
				\end{aligned}
			\end {cases}.
		\end{align}

		The transmitted signal $x(t)$ is then obtained by multiplexing the truncated/concatenated single-numerology signals associated with both the numerologies. The mixed signal takes different forms for numerology 1 and 2 with respect to their own symbol duration.  Specifically, the signal $x(t)$ on the time duration $[kT^{(1)}, (k+1)T^{(1)}]$  can be expressed as
		\begin{align} \label{eq_tx_sig}
			x_k(t)= x^{(1)}_k(t) + x^{(2)}_{\lfloor \frac k v \rfloor}(t),
		\end{align}
		where $\lfloor \cdot \rfloor$ is the floor function. The signal $x(t)$ on $[kT^{(2)}, (k+1)T^{(2)}]$ can be written as
		\begin{align} \label{eq_tx_sig_2}
		 	x_k(t) = \sum_{q=0}^{\nu-1}x^{(1)}_{\nu k+q} + x^{(2)}_k(t).
		\end{align}

		Eq. \eqref{eq_tx_sig} and \eqref{eq_tx_sig_2} mathematically describe  what Fig. \ref{fig:timefreqillustration} illustrates, i.e, how signals of different symbol duration multiplex in the time domain.  In particular, the $k$-th symbol associated with numerology $1$ overlaps with a truncated portion of the $\lfloor \frac {k}{\nu}\rfloor$-th symbol associated with numerology $2$, and  the $k$th symbol associated with numerology $2$ coincides with a block of ${\nu}$  consecutive symbols starting from the $({\nu}k)$-th symbol associated with numerology $1$.

		The received signal is passed through SN-OFDM demodulators in order to separate signals on different subcarriers within the numerology and reject signals from the other numerology.  In this paper, we focus on investigating the intrinsic interference caused by the MN-OFDM modulator/demodulator only, while the extrinsic factors such as the channel and noise  are not considered.  Thus, the received signal $r(t) \!=\! x(t)$.  The recovered complex symbol transmitted in the time duration $[kT^{(i)}, (k+1)T^{(i)}]$ over the $m$-th subcarrier associated with numerology $i$ can be obtained as
		\begin{align}\label{eq:rx_est_sig}
			\tilde{s}^{(i)}_{k,m}  =\int_{kT^{(i)}}^{(k+1)T^{(i)}}x_{k}(t) {\phi^{(i)}_m}^*(t-kT^{(i)}) \ dt.
		\end{align}

	\section{Characterizing the INI in MN-OFDM systems } \label{sec:cha_INI}
		
		Two signals are orthogonal to each other if their inner product is zero \cite{vetterli1995wavelets}, otherwise they are correlated, and interfere with each other. Thus, the interference between two MN subcarriers can be characterized by their inner product with the magnitude as an indicator of the level of correlation. For a single numerology, the set of SN subcarriers $\{\phi^{(i)}_m(t)\}, m =0,1,\cdots \}$  forms an orthonormal basis on the interval $[0,T^{(i)}]$,  and the subcarriers are non-correlated and orthogonal to each other\cite{vetterli1995wavelets}. As such, we will only focus on mixed-numerology in this section, where the closed-form formulas for INI between subcarriers will be derived for continuous-time and discrete-time signals, respectively.
		%An important characteristic of  orthogonal/uncorrelated signals is that,  when they are added together,  they do not interfere with each other and can be perfectly separated. What makes

		\subsection{Continuous-time MN signal} \label{subsec:con_sig}

			Let $\phi^{(1)}_m(t)$ and $\phi^{(2)}_n(t)$ be subcarrier $m$ associated with numerology $1$  and subcarrier $n$ associated with numerology $2$, respectively. The corresponding inner product  can be obtained as
			\begin{align}\label{eq:correlation_2}
				\rho^{(1\leftarrow 2)}_{m,n}&=
				\langle\phi^{(1)}_m(t),\phi^{(2)}_n(t)\rangle = \int_{-\infty}^{\infty} {\phi^{(1)}_m}^*(t) {\phi^{(2)}_n}(t)dt \nonumber\\
				& =\frac 1 {\sqrt{T^{(1)}T^{(2)} }} \int_{0}^{T^{(1)} }\exp\left(-j2\pi(\frac{m}{T^{(1)}}- \frac{n}{T^{(2)}}) t\right ) dt\nonumber\\
				& = \frac 1 {\sqrt{T^{(1)}T^{(2)} }} \exp\left(-j2\pi (\frac{m}{T^{(1)}}- \frac{n}{T^{(2)}}) \frac {T^{(1)}}{2}\right )  \nonumber\\
				& \qquad \qquad \quad \int_{-\frac {T^{(1)}}{2}}^{\frac {T^{(1)}}{2}} \exp\left(-j2\pi(\frac{m}{T^{(1)}}- \frac{n}{T^{(2)}}) t\right ) dt\nonumber\\
				& = \sqrt{\frac 1 {\nu }} \exp\left(-j\pi d \right)  \text{sinc}\left (d\right ),
			\end{align}
			where $\langle\cdot \rangle$ is the inner product function, $\text{sinc}(x) = {\sin(\pi x)}/{(\pi x)}, \, x \in \mathbf R$, and $d$ refers to the relative distance between subcarrier $m$ and subcarrier $n$ expressed as
			\begin{align} \label{eq:reldis}
				d &=(\frac{m}{T^{(1)}}- \frac{n}{T^{(2)}})T^{(1)}= \frac{m\Delta f^{(1)} -n \Delta f^{(2)}}{\Delta f^{(1)}}\nonumber\\
				& ={m} -  \frac {n}{\nu},
			\end{align}
			which is the actual distance between the two subcarrier centre normalized by $\Delta f^{(1)}$. Swapping the order of two subcarriers turns the corresponding inner product into the complex conjugate of the original one as
			\begin{align}
				\rho^{(1\leftarrow 2)}_{m,n}={\rho^{(2\leftarrow 1)}_{n,m} }^*
			\end{align}
			and their common magnitude is written as
			\begin{align}\label{eq:correlation_amp}
				\left \vert \rho^{(1\leftarrow 2)}_{m,n}\right \vert = \left \vert \rho^{(2\leftarrow 1)}_{n,m}\right \vert  = \sqrt{\frac 1 {\nu }}  \left\vert \text{sinc}\left (d \right )  \right \vert  
			\end{align}
			where $\vert \cdot\vert$  is the absolute operator. This suggests that the interference between any two subcarriers associated with different numerologies is mutual  and equal in magnitude.

			\begin{figure}[t]
				\centering
				\includegraphics[width=1\linewidth]{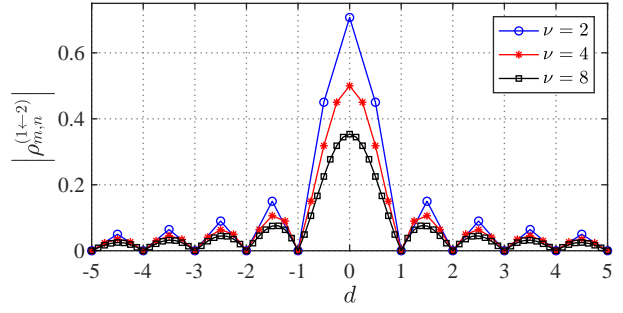}
				\caption{Magnitude $\rho^{(1\leftarrow 2)}_{m,n}$ in \eqref{eq:correlation_amp} vs relative distance $d$ in \eqref{eq:reldis} for continuous-time signals.}
				\label{fig:sim8}
			\end{figure}

			It is clear from \eqref{eq:correlation_2} that the sinc function, which originates from the rectangular pulses, is the dominant term of the inner product. Eq. \eqref{eq:correlation_amp} shows that the magnitude of the inner product is dependent on the relative distance $d$ which is determined by the subcarrier indices $m$ and $n$ and the numerology scaling factor $\nu$. The magnitude can be expected to decrease as the relative distance increases, and a bigger scaling factor leads to a smaller magnitude. Fig. \ref{fig:sim8} depicts the magnitude of the inner product of subcarriers associated with different numerologies related to each other with the scaling factor $\nu\in\{2,4,8\}$. It can be seen in the figure that the magnitude is inversely proportional to $\nu$, which indicates that a large scaling factor implies less correlated subcarriers. Thus, numerologies associated with bigger subcarrier spacing are more resilient to INI. In addition, the magnitude appears to decrease with the increase in the relative distance,  which shows that the interference from close subcarriers is higher than that from relatively distant ones.

		\subsection{Discrete-time MN signal}

			Assume a common sampling duration $T_s = {T^{(i)}}/{N^{(i)}}$ for both numerologies. Then the $l$-th sample of the signal in \eqref{eq:sn_sig} can be expressed as
			\begin{align}
				x_{k}^{(i)}[l] = x_{k}^{(i)}\left (lT_s\right ) = \sum_{m=1}^{N^{(i)}}s^{(i)}_{k,m} \phi^{(i)}_m[l],
			\end{align}
			where
			\begin{align}
				\phi^{(i)}_m[l] =\frac{\exp(j2\pi \frac{ml}{N^{(i)}})}{\sqrt{N^{(i)}}} , l = 0,1,\cdots, N^{(i)} -1.
			\end{align}
			This is the inverse Discrete Fourier Transform (DFT) of the transmitted signal, which is much easier to implement by performing inverse fast Fourier transform (FFT) with integrated circuits.
			
			The inner product of the discrete MN subcarriers  is derived as
			\begin{align}\label{eq:correlation_dis}
				\bar \rho^{(1\leftarrow 2)}_{m,n}&=
				<\phi^{(1)}_m(t),\phi^{(2)}_n(t)> \nonumber\\&= \frac 1 {\sqrt{N^{(1)}N^{(2)}} }\sum_{l=0}^{N^{(1)}}
				\exp \left ( j2\pi \left (\frac m {N^{(1)}} - \frac n {N^{(2)}}\right)  l\right) \nonumber\\
				& \overset{(a)}{=}    \frac 1 {\sqrt{N^{(1)}N^{(2)}} } \frac{1- \exp(j2\pi d)}{1 - \exp(j 2\pi \frac d {N^{(1)}})} \nonumber \\
				& = \frac 1 {\sqrt{\nu}}  \exp(j \pi \frac{N^{(i)}-1}{N^{(i)}}d) \frac{\text{sinc}(d)}{\text{sinc} (\frac d {N^{(1)}})},
			\end{align}
			where equality $(a)$ follows that $\sum_{n=0}^{N-1} x^n = \frac {1-x^N}{1-x}$, $x \in \mathbf R$. The magnitude in this case is
			\begin{align} \label{eq:mag_discrete}
				\left \vert \bar \rho^{(1\leftarrow 2)}_{m,n} \right \vert =  \sqrt{\frac 1 {\nu }} \frac{ \left\vert \text{sinc}\left (d \right )  \right \vert }{\left\vert \text{sinc}\left (\frac d {N^{(1)}} \right )  \right \vert}.
			\end{align}
			
			\begin{figure}
				\centering
				\includegraphics[width=1\linewidth]{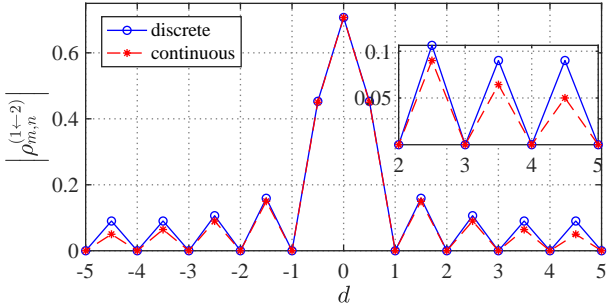}
				\caption{Magnitude $\rho^{(1\leftarrow 2)}_{m,n}$ vs relative distance $d$ for continuous-time and discrete-time signals with number of samples $N^{(1)} = 8$, $\nu =2$.}
				\label{fig:sim9}
			\end{figure}
			
			Since the subcarriers in SN systems  are orthogonal to each other, discretization has impact on interference, and thus the two approaches (continuous-time and discrete-time) are equivalent. This is not the case in MN systems. Comparing \eqref{eq:mag_discrete} with \eqref{eq:correlation_amp}, it can be seen that the magnitude of the interference in the discrete-time systems differs from the continuous-time systems by a factor. Fig. \ref{fig:sim9} compares the magnitude of the inner product between discrete-time and continuous-time signals where both curves show similar varying trend, i.e., they decay with distance. However, curves for the discrete-time case decays relatively slower. The impact of discretization with different sampling rates on the INI will be studied in the next subsection.

		\subsection{The impact of the discretization on the INI}
			Physical signals are usually defined in continuous time, but signal processing is conducted more efficiently digitally and for discrete-time signals.
			In theory, when the number of samples in a discrete-signal approaches infinity, the inner product of MN subcarriers becomes equal to that of the continuous case, i.e.,
			\begin{align}
				\lim_{N^{(1)} \rightarrow \infty} \bar \rho^{(1\leftarrow 2)}_{m,n}   = \sqrt{\frac 1 {\nu }} \exp\left( j\pi d \right)  \text{sinc}\left (d\right )= \rho^{(1\leftarrow 2)}_{m,n} 
			\end{align}

			In practice, for any finite value of  $N^{(1)}$, the magnitude of the interference between the MN subcarriers $m$ and $n$ for discrete signals differs from the continuous ones by a factor of 
			\begin{align} \label{eq:fact}
				\beta =    \frac{1}{\vert\text{sinc} (\frac d {N^{(1)}})\vert}.
			\end{align}

			Eq.\eqref{eq:fact} implies that the discrete subcarriers suffer more interference compared to its continuous counterpart giving that $\beta  >1, \forall d  \neq 0$. This can also be explained by the fact that the sampled subcarriers appear closer to each other than the continuous ones in the signal subspace due to the finite number of representative samples.

\begin{figure}
				\centering
				\includegraphics[width=1\linewidth]{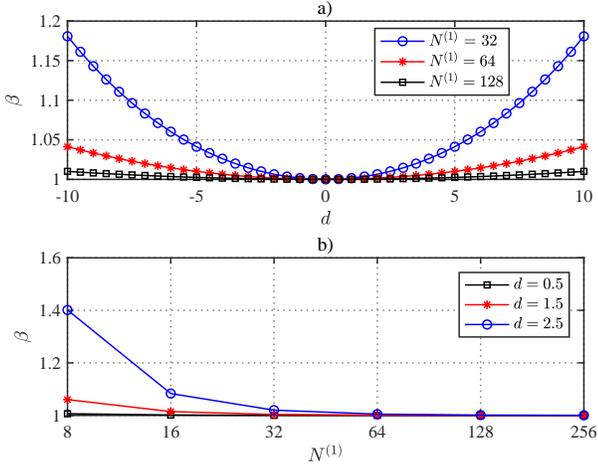}
				\caption{Factor $\beta$ in \eqref{eq:fact} with two numerologies with  $\Delta f^{(1)} =  2\Delta f^{(2)}$. a) $\beta$ vs. relative distance $d$. b) $\beta$ vs. number of samples $N^{(1)}$. }
				\label{fig:disconfactor}
			\end{figure}

			The factor in \eqref{eq:fact} implies that the level of influence of discretization increases as the absolute value of relative distance $d$ increases and/or the number of samples $N^{(i)}$ decreases. Thus, over-sampling signals could improve system error performance in discrete-time MN systems. In addition, the INI contribution from the subcarriers that are relatively distant is affected more by discretization than from those closer subcarriers, as a larger $d$ leads to a greater $\beta$.  Fig. \ref{fig:disconfactor} depicts the effects of discretization on the INI in terms of the absolute value of factor $\beta$ in \eqref{eq:fact} evaluated for different subcarrier distances and the number of samples $N^{(i)}$.  It can be seen that the impact changes less rapidly with regard to the relative distance $d$ for a higher value of $N^{(i)}$ and that the impact becomes barely visible with $N^{(i)} \geq 64$, and $d \leq 2.5$.
			
			Based on the analysis in Subsection \ref{subsec:con_sig}, the level of the interference of a subcarrier in one numerology is primarily determined by those interfering subcarriers that are close to the interfered subcarrier. Thus, we can conclude that the INI for discrete-time signals can be used to closely approximate the INI for continuous-time signals  with an error of $( \beta  -1)\%$. Given a predefined tolerance error, the minimum sampling rate can be obtained accordingly using \eqref{eq:fact}.
			
			\subsection{Subsets of orthogonal MN subcarriers}
				The zero-crossings of the sinc functions in \eqref{eq:correlation_2} and \eqref{eq:correlation_dis} for the continuous-time and discrete-time system, respectively, indicate that there are some interference-free subcarriers despite the overall loss of orthogonality in MN-OFDM systems. Finding these subcarriers is important, as they can be used as pilot subcarriers for a better channel estimation, or allocated to users experiencing worse channel conditions to improve fairness.
				
				%Two complex signals are orthogonal over an certain interval if their inner product is zero over the interval\cite{vetterli1995wavelets}, otherwise they interfere with each other. 
				The inner product of any two subcarriers associated with the same numerology is zero, thus, subcarriers within a numerology are orthogonal to each other. For subcarriers associated with different numerologies, solving the equation 
				\begin{align}
				\begin{cases}
					\left \vert \rho^{(1\leftarrow 2)}_{n,m}\right \vert  =0 \\\\
					\left \vert \bar \rho^{(1\leftarrow 2)}_{n,m}\right \vert  =0 \\
				\end{cases}
				\Rightarrow  \sqrt{\frac 1 {\nu }}  \left\vert \text{sinc}\left ({m} -  \frac{n} {\nu}\right )  \right \vert  =0,
				\end{align} 
				enables us to establish the condition for the orthogonality of MN subcarriers for both the discrete and continuous signals as, %which are and not entered at the same frequency and, as
				\begin{align}\label{eq:orthog_con_1}
				\left ({m} -  \frac{n} {\nu}\right ) \in \mathbf Z  \ \equiv \ \frac{n} {\nu} \in \mathbf Z,
				\end{align}
				where subcarrier $m$, $n$ are associated with the numerology with a greater subcarrier spacing and a smaller one, respectively.
				
				It is clear  from \eqref{eq:orthog_con_1} that the orthogonality between two MN subcarriers is only determined by the index of the subcarrier with a smaller subcarrier spacing. More specifically, the two subcarriers are orthogonal to each other if the index of the subcarrier with a smaller subcarrier spacing is an integer multiple of the subcarrier spacing ratio $\nu$,  regardless of the index of the other subcarrier. %On the contrary,  if the index of the subcarrier with a smaller subcarrier spacing  is not integer multiple of the subcarrier spacing ratio $\nu$, it will interfere with the other subcarrier. Thus, for two numerologies, there exist a subset of  interference-free subcarriers. 
				As an example for a two-numerology systems with $\nu =2$, there are three orthogonal subsets of subcarriers: the subset of all the subcarriers associated with numerology 1, the subset of all the subcarriers associated with numerology 2, the subset of MN subcarriers which includes all the subcarriers associated with numerology 1 and the even indexed subcarriers associated with numerology 2.
				
				\begin{figure}[t]
					\centering
					\includegraphics[width=0.9\linewidth]{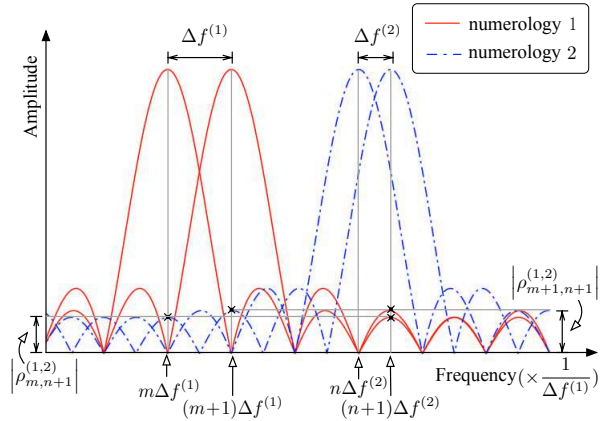}
					\caption{Illustration of the INI between subcarriers associated with two numerologies with  $\Delta f^{(1)} =  2\Delta f^{(2)}$ over the time interval $[0, T^{(1)}]$, where $T^{(1)} = 1 /{\Delta f^{(1)}}$. Note: The figure is not to show the orthogonality between subcarriers associated with same numerology because a common symbol duration ($T^{(1)}$) is used for both numerologies.}
					\label{fig:mixnumfreq}
				\end{figure}

				Fig. \ref{fig:mixnumfreq} illustrates some orthogonal MN subcarriers for two numerologies with $\nu=2$. Two adjacent subcarriers indexed by $m$ and $m+1$ associated with numerology $1$,  and  two adjacent subcarriers indexed by $n$ and $n+1$ are associated with numerology $2$ ($n$ is multiple of 2) are shown in the figure.  The subcarrier $n$ is orthogonal to both subcarriers $m$ and $m+1$ with  $\rho^{(1\leftarrow 2)}_{m,n} = \rho^{(1\leftarrow 2)}_{m+1,n} =0$.  The subcarrier $n+1$ is not orthogonal to subcarrier $m$ or to subcarrier $m+1$ with $\left \vert \rho^{(1\leftarrow 2)}_{m,n+1} \right \vert =  \left \vert \rho^{(2\leftarrow 1)}_{n+1,m} \right \vert > 0$, $\left \vert\rho^{(1\leftarrow 2)}_{m+1,n+1}\right \vert = \left \vert\rho^{(2\leftarrow 1 )}_{n+1,m+1}\right \vert  > 0$.
			%This can be confirmed by \eqref{eq:mag_discrete} and \eqref{eq:mag_con} as 
			%\begin{align} 
				%\left \vert \bar \rho^{(1\leftarrow 2)}_{m,n} \right \vert =  \sqrt{\frac 1 {\nu }} \frac{ \left\vert \text{sinc}\left (d \right )  \right \vert }{\left\vert \text{sinc}\left (\frac d {N^{(1)}} \right )  \right \vert} = \frac{\left \vert  \rho^{(1\leftarrow 2)}_{m,n} \right \vert}{\left\vert \text{sinc}\left (\frac d {N^{(1)}} \right )  \right \vert} \geq \left \vert  \rho^{(1\leftarrow 2)}_{m,n} \right \vert,	
			%\end{align}
			%where  the quality holds when $d \in \mathbf  Z$, $\left \vert \bar \rho^{(1\leftarrow 2)}_{m,n} \right \vert = \left \vert \bar \rho^{(1\leftarrow 2)}_{m,n} \right \vert =0$,  otherwise,  the inequity holds with $\vert \text{sinc}\left (\frac d {N^{(1)}} \right )  \vert< 1$.According to \eqref{eq:mag_discrete} and \eqref{eq:mag_con}}, it is obvious that the inner products for discrete signals and continuous signals are different in magnitude.
			%\begin{align}
				%\Delta = 	\left \vert \bar \rho^{(1\leftarrow 2)}_{m,n} \right \vert - \left \vert  \rho^{(1\leftarrow 2)}_{m,n} \right \vert 
			%\end{align}
			
			%Specifically, the former is greater than latter with nonzero relative distance due to the  denominator \eqref{eq:mag_discrete}  is less than 1,With the number of samples $N^{(1)}$  increases, the difference becomes smaller. 

	\section{Conclusions} \label{sec:con}
	
	In this paper, the interference between subcarriers associated with different numerologies was studied to understand MN-OFDM systems' behavior. Novel reduced-form formulas for the INI for continuous-time and discrete-time signals were derived, which reveal the influencing factors more explicitly compared to traditional SoE forms.  Based on the derived reduced-form formulas, the impact of discretization on the INI was investigated, and it was shown that subcarriers suffer more interference in discrete-time systems. Moreover, orthogonality within subsets of subcarriers in mixed-numerology systems was discussed, which provides a guidance on pilot subcarrier allocation and user scheduling. This study emphasized the influencing factors of INI contributions, and provided a valuable reference and useful guidance for the design of future MN-OFDM systems with effective INI mitigation strategies.

	\ifCLASSOPTIONcaptionsoff
	\newpage
	\fi
	
	\bibliographystyle{IEEEtran}
	\bibliography{ms.bib}

\begin{thebibliography}{10}
\providecommand{\url}[1]{#1}
\csname url@samestyle\endcsname
\providecommand{\newblock}{\relax}
\providecommand{\bibinfo}[2]{#2}
\providecommand{\BIBentrySTDinterwordspacing}{\spaceskip=0pt\relax}
\providecommand{\BIBentryALTinterwordstretchfactor}{4}
\providecommand{\BIBentryALTinterwordspacing}{\spaceskip=\fontdimen2\font plus
\BIBentryALTinterwordstretchfactor\fontdimen3\font minus
  \fontdimen4\font\relax}
\providecommand{\BIBforeignlanguage}[2]{{%
\expandafter\ifx\csname l@#1\endcsname\relax
\typeout{** WARNING: IEEEtran.bst: No hyphenation pattern has been}%
\typeout{** loaded for the language `#1'. Using the pattern for}%
\typeout{** the default language instead.}%
\else
\language=\csname l@#1\endcsname
\fi
#2}}
\providecommand{\BIBdecl}{\relax}
\BIBdecl

\bibitem{series2015imt}
M.~Series, ``{IMT Vision--Framework and Overall Objectives of the Future
  Development of IMT for 2020 and Beyond},'' \emph{Recommendation ITU}, 2015.

\bibitem{yang2020mixed}
B.~Yang, L.~Zhang, O.~Onireti, P.~Xiao, M.~A. Imran, and R.~Tafazolli,
  ``{Mixed-Numerology Signals Transmission and Interference Cancellation for
  Radio Access Network Slicing},'' \emph{IEEE Transactions on Wireless
  Communications}, pp. 1--14, 2020.

\bibitem{zhang2017multi}
L.~Zhang, A.~Ijaz, P.~Xiao, and R.~Tafazolli, ``{Multi-Service System: An
  Enabler of Flexible 5G Air Interface},'' \emph{IEEE Commun. Mag}, vol.~55,
  no.~10, pp. 152--159, 2017.

\bibitem{zaidi2016waveform}
A.~A. Zaidi, R.~Baldemair, H.~Tullberg \emph{et~al.}, ``{Waveform and
  Numerology to Support 5G Services and Requirements},'' \emph{IEEE Commun.
  Mag.}, vol.~54, no.~11, pp. 90--98, 2016.

\bibitem{zhang2017subband}
L.~Zhang, A.~Ijaz, P.~Xiao \emph{et~al.}, ``{Subband Filtered Multi-Carrier
  Systems for Multi-Service Wireless Communications},'' \emph{IEEE Trans.
  Wireless Commun.}, vol.~16, no.~3, pp. 1893--1907, 2017.

\bibitem{yli2017efficient}
J.~Yli-Kaakinen, T.~Levanen, S.~Valkonen, K.~Pajukoski, J.~Pirskanen,
  M.~Renfors, and M.~Valkama, ``{Efficient Fast-Convolution-Based Waveform
  Processing for 5G Physical Layer},'' \emph{IEEE J. Select. Areas Commun.},
  vol.~35, no.~6, pp. 1309--1326, 2017.

\bibitem{zhang2018mixed}
X.~Zhang, L.~Zhang, P.~Xiao \emph{et~al.}, ``{Mixed Numerologies Interference
  Analysis and Inter-Numerology Interference Cancellation for Windowed OFDM
  Systems},'' \emph{IEEE Trans. Veh. Technol.}, 2018.

\bibitem{mao2020interference}
J.~Mao, L.~Zhang, P.~Xiao, and K.~Nikitopoulos, ``{Interference Analysis and
  Power Allocation in the Presence of Mixed Numerologies},'' \emph{IEEE Trans.
  Wireless Commun.}, 2020.

\bibitem{3gpp_38_211}
3GPP, ``{NR; Physical Channels and Modulation},'' {3rd Generation Partnership
  Project}, TS 38.211, 09 2018.

\bibitem{vetterli1995wavelets}
M.~Vetterli and J.~Kovacevic, \emph{{Wavelets and Subband Coding}}.\hskip 1em
  plus 0.5em minus 0.4em\relax Prentice-hall, 1995.

\end{thebibliography}

\end{document}